\documentclass[prb,twocolumn,floatfix,showpacs]{revtex4}

\usepackage{amsmath}
\usepackage{graphics}

\newcommand{\vect}[1]{\boldsymbol{#1}}
\newcommand{\aop}{\hat{a}}
\newcommand{\figwidth}{7.75cm}

\begin{document}

\title{Universal quantum computing with correlated spin-charge states}

\author{Jordan Kyriakidis}
\affiliation{Department of Physics and Atmospheric Science, Dalhousie
  University, Halifax, Nova Scotia, Canada, B3H 3J5}

\author{Guido Burkard}
\affiliation{Department of Physics and Astronomy, University of Basel,
  Klingelbergstrasse 82, CH-4056 Basel, Switzerland}

\date{\today}

\begin{abstract}
  We propose a universal quantum computing scheme in which the
  orthogonal qubit states $|0\rangle$ and $|1\rangle$ are identical in
  their single-particle spin and charge properties.  Each qubit is
  contained in a single quantum dot and gate operations are induced
  all-electrically by changes in the confinement potential.  Within
  the computational space, these qubits are robust against
  environmental influences that couple to the system through
  single-particle channels.  Due to the identical spin and charge
  properties of the $|0\rangle$, $|1\rangle$ states, the lowest-order
  relaxation and decoherence rates $1/T_1$ and $1/T_2$, within the
  Born-Markov approximation, both vanish for a large class of
  environmental couplings.  We give explicit pulse sequences for a
  universal set of gates (phase, $\pi/8$, Hadamard, \textsc{cnot}) and
  discuss state preparation, manipulation, and detection.
\end{abstract}

% PACS:
% 03.67.Lx Quantum computation
% 73.21.La Quantum dots
% 85.35.Be Quantum well devices (quantum dots, quantum wires, etc)

\maketitle

\section{Introduction}
\label{sec:introduction}

Proposals for quantum computing architectures based on semiconductor
devices\cite{Loss98:Quantum-computation, Burkard99:Coupled-quantum,
  kane98:silic.based.nuclear, Troiani00:Exploiting-exciton-exciton,
  Biolatti00:Quantum-Information, hayas03:coher.manip.elect,
  li03:all.optic.quant} are attractive for their scalability; once the
few-qubit problem is solved, massive scalability is not expected to
pose insurmountable barriers either in resource requirements or
fabrication precision.  This is primarily due to the sustained and
continued improvements in epitaxy and lithography over the past few
decades, and the ability with which new techniques, often developed in
industry, are transferred to basic research laboratories.  On the
other hand, semiconductor environments are hardly systems of pristine
quality and isolation, and there are severe trade-offs between long
coherence times and short access times.

Pure spin qubits, for example, couple relatively weakly to their
environment.\cite{Petta05:Coherent-Manipulation-of-Coupled}  Their
dipole tails are often negligibly weak and spin exchange effects,
while potentially strong, are short range.  But precisely because of
this weak environmental coupling, spin qubits may be potentially
difficult to control and manipulate.  For single-particle qubits,
local Zeeman tuning is required to rotate bits.  The opposite scenario
is often true for charge qubits.  Here, control may be attained very
quickly with metallic gates or
optics.\cite{hayas03:coher.manip.elect,li03:all.optic.quant}  However
relaxation and decoherence times can be very fast, requiring even
faster switching times.

This begs the question of whether there exist hybrid qubits which
accentuate the positives and mitigate the negatives.  We show below that
this does indeed seem the case if a single qubit is judiciously
defined as a correlated few-body system whose charge and spin degrees
of freedom are entangled.  These strong correlations should
additionally be effective at suppressing relaxation and decoherence
through single-particle channels.  Indeed our two orthogonal qubit
states are identical in their single-particle spin and charge degrees
of freedom; differences only show up in their two-body correlation
functions.

Sources of decoherence and dissipation can also be broadly classified
as spin based or charge based.  Both destroy the unitary dynamics of
the system either by taking it outside the computational subspace, or
by remaining within the computational subspace, but causing either
uncontrolled qubit flips, or pure dephasing without dissipation of
energy.  This will generally occur whenever an environmental influence
couples differently to each qubit state.  For example, if the two
qubit states differ in their spin, then random magnetic fields are an
issue.  For single-particle qubits, this will always be the case, and
likewise for two-particle qubits; it is not possible to define two
orthogonal one- or two-particle states with identical spin and charge
densities.  A three particle system, however, can be constructed in
which \emph{both} the charge density and the spin of the two
orthogonal $|0\rangle$ and $|1\rangle$ states are identical.

We show that the qubits we define below admit a universal set of one
and two-qubit gates, and we give explicit gate pulse sequences which
implement this universal set.  We also discuss issues of decoherence
and relaxation among the qubits and show that, for a broad class of
environments, including certain spin dependent ones, relaxation and
dephasing are absent ($1/T_1 = 1/T_{\varphi} = 0$) within the
lowest-order Born-Markov approximation.  We expect the residual
decoherence rate due to higher-order couplings, non-Markovian effects,
and other, weakly coupled, environments to be small.  We also discuss
extensions to the model of system-environment coupling, and comment on
issues of state preparation and detection.

In the following section, we describe our model electronic Hamiltonian
consisting of two many-body parabolic-elliptic quantum dots, with
long-range intradot Coulomb repulsion.  In
Sec.~\ref{sec:qubit-construction} we construct our qubits and
demonstrate how correlations produce orthogonal $|0\rangle$ and
$|1\rangle$ states with identical spin and single-particle charge
densities.  Section~\ref{sec:univ-quant-logic} contains explicit
implementations---in the form of pulse sequences---for a universal set
of quantum logic gates (Hadamard, $\pi/8$, Phase, and \textsc{cnot}
gates).  Section~\ref{sec:decoh-diss} demonstrates that, to lowest
order, intra-qubit relaxation and dephasing is absent for all pure spin
and pure charge environments which couple to the qubit through
single-particle channels.  Finally, in Sec.~\ref{sec:init-meas}, we
briefly discuss issues of state preparation and detection.

\section{Model Hamiltonian}
\label{sec:model-hamiltonian}

We consider two coupled dots with Hamiltonian
\begin{equation}
  \label{eq:hamil}
  \hat{H} = \hat{H}_{\text{dot1}} + \hat{H}_{\text{dot2}} +
  \hat{H}_{\text{coupl}},
\end{equation}
where the first two terms denote the individual quantum dots whereas
the third denotes interdot coupling.  For $\hat{H}_{\text{coupl}}$, we
shall take a simple coupling Hamiltonian but, within each quantum dot,
we shall take full long-range repulsive interactions into account
(exactly).  We first focus on a single qubit and subsequently discuss
two-qubit interactions.

We encode a single qubit in a single elliptically
confined\cite{ellipNote} two-dimensional lateral quantum dot.  We
place three interacting electrons in the dot and consider the
two-dimensional subspace spanned by the $S = 1/2, S_z = -1/2$ spin
sector.  Single qubit rotations are created by tuning the eccentricity
of the elliptic confinement
potential,\cite{Kyriakidis05:Coherent-rotations} whereas two-qubit
operations, as we show below, are created by controlling the coupling
between two adjacent quantum
dots.\cite{Loss98:Quantum-computation,Burkard99:Coupled-quantum}

The Hamiltonian of a single dot is given by $\hat{H}_{\text{dot1}} =
\hat{H}_{\text{1body}} + \hat{H}_{\text{Coul}}$, where
\begin{equation}
  \label{eq:ellipHamil}
  \hat{H}_{\text{1body}} = \frac{1}{2m} \left( \hat{\vect{p}} - \frac{e}{c}
    \hat{\vect{A}} \right)^2 +
  \frac{1}{2} m \left( \omega_x^2 \hat x^2 + \omega_y^2 \hat y^2 \right).
\end{equation}
We take a magnetic field $\vect{B} = (0, 0, B)$ perpendicular to the
plane of the dot.  The Hamiltonian~(\ref{eq:ellipHamil}) can be
exactly diagonalized with canonical Bose operators
$\hat{a}_1^{\dagger}$, $\hat{a}_2^{\dagger}$ and their Hermitian
conjugates
as\cite{Kyriakidis05:Coherent-rotations,madhav94:elect.proper.anisot}
\begin{equation}
  \label{eq:singDotHamil}
  \hat{H}_{\text{1body}} = 
  \hbar \Omega_+ \left(\aop_1^{\dag}\aop_1 + \frac{1}{2}\right) + 
  \hbar \Omega_- \left(\aop_2^{\dag}\aop_2 + \frac{1}{2}\right),
\end{equation}
where
\begin{subequations}
  \begin{gather}
    \Omega_\pm = \frac{1}{2}\sqrt{\omega_0^2 + \omega_c^2 \pm 2
      \Omega^2}, \qquad \omega_{c} = \frac{eB}{mc},\\
    \omega_0 = \sqrt{\omega_c^2 + 2 \left(\omega_x^2 + \omega_y^2
      \right)}, \quad 
    \Omega = \left(\omega_-^4 + \omega_c^2\omega_0^2\right)^{1/4},\\
    \omega_- = \sqrt{\omega_x^2 - \omega_y^2}.
  \end{gather}
\end{subequations}
We build many-body states through antisymmetrized products of
single-particle states $|n m\rangle$, where $\aop_1^{\dag} \aop_1 |n
m\rangle = n |n m\rangle$ and $\aop_2^{\dag} \aop_2 |n m\rangle = m |n
m\rangle$.

The long-range Coulomb interaction $\hat{H}_{\text{Coul}}$ can then be
written in the usual second-quantized form as 
\begin{equation}
  \label{eq:HCoul}
  \hat{H}_{\text{Coul}} = \frac{1}{2} \sum V_{ijkl}
  c^{\dag}_{i \sigma} c^{\dag}_{j \sigma'} c_{l \sigma'} c_{k \sigma},
\end{equation}
where all indices are summed over.  An explicit and exact closed form
expression for the matrix element
\begin{equation}
  \label{eq:VMatel}
  V_{ijkl} = \int\!d^2q\, \frac{e^2}{2 \pi q \varepsilon}
  (m_i n_i, m_j n_j| 
  \text{e}^{i \vect{q} \cdot (\hat{\vect{r}}_1 - \hat{\vect{r}}_2)}
  |m_k n_k, m_l n_l),
\end{equation}
is derived in Ref.~[\onlinecite{Kyriakidis05:Coherent-rotations}].

\section{Qubit Construction}
\label{sec:qubit-construction}

For definiteness, we consider three singly-occupied orbitals $|n
m\rangle = |0 0\rangle,\ |0 1\rangle,\ |0 2\rangle$ corresponding to
the three lowest-energy orbitals in the lowest Landau level.  With
this orbital occupation, the $S=1/2,\ S_z=-1/2$ subspace is
two-dimensional and is spanned by our orthogonal qubit states
\begin{subequations}
  \label{eq:qubits}
  \begin{gather}
    \label{eq:qbit0}
    |0\rangle \equiv \frac{1}{\sqrt{6}} 
    \bigl(
      2|\downarrow \downarrow \uparrow  \rangle -
       |\downarrow \uparrow   \downarrow\rangle -
       |\uparrow   \downarrow \downarrow\rangle 
    \bigr),
    \\
    \label{eq:qbit1}
    |1\rangle \equiv \frac{1}{\sqrt{2}}
    \bigl(
      |\downarrow \uparrow   \downarrow\rangle -
      |\uparrow   \downarrow \downarrow\rangle
    \bigr).
  \end{gather}
\end{subequations}
Each term on the right is a single antisymmetrized state: $|s_0 s_1 s_2\rangle
\equiv c_{0 0 s_0}^{\dagger} c_{0 1 s_1}^{\dagger} c_{0 2
  s_2}^{\dagger} |\text{vacuum}\rangle$, where the operator $c_{n m
  s}^{\dagger}$ creates an electron in state $|n m s\rangle$.  The
spin configurations in Eq.~(\ref{eq:qubits}) are 
(up to an overall exchange of spin up and down) those in
Ref.~[\onlinecite{Divincenzo00:Universal-quantum}]; however, the 
electron states differ in their orbital degrees of freedom.  In
particular, the states~(\ref{eq:qubits}) have particles sitting in
orthogonal orbitals; this orthogonality is required for the charge
densities to be identical during gate operations.

\begin{figure}
  \resizebox{\figwidth}{!}{\includegraphics*{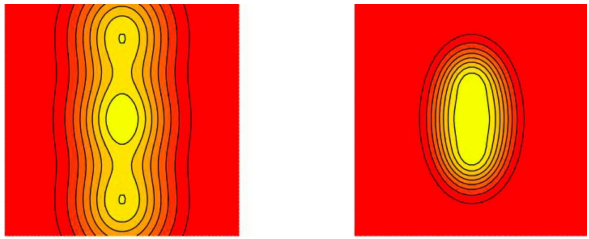}}
  \caption{\label{fig:charge-dens}(Color online) Identical charge
    density $\langle \rho(\vect{r}) \rangle$ for both qubit states.
    Both plots have $\omega_y / \omega_x = 1/2$.  The left plot is at
    zero magnetic field whereas the right has $\omega_c / \omega_x =
    5$.  For $\omega_x = 1$~meV, this corresponds to $B_z \approx 3$~T
    for GaAs.}
\end{figure}

The states in Eq.~(\ref{eq:qubits}) cannot be written as single Slater
determinants in \emph{any} single-particle basis; they are correlated
states with entangled spin and charge degrees of freedom.  These
correlations enable the states to be both orthogonal to each other and
yet exhibit identical single particle properties.  Both qubit states
in (\ref{eq:qubits}) have spin $S = 1/2,\ S_z = -1/2$.  Furthermore,
defining the charge density operator as $\hat{\rho}(\vect{r}) = \sum_i
\delta(\vect{r} - \hat{\vect{r}}_i)$, we find $\langle \rho
\rangle_{|0\rangle} = \langle \rho \rangle_{|1\rangle} = \sum_i
|\psi_{0i}(\vect{r})|^2$, where $\psi_{nm}(\vect{r}) \equiv
\langle\vect{r}|n m\rangle$ is a real-space eigenstate of
Eq.~(\ref{eq:ellipHamil}).  The density is plotted in
Fig.~\ref{fig:charge-dens} for two different values of $z = \omega_c /
\omega_x$.

\begin{figure}
  \resizebox{\figwidth}{!}{\includegraphics*{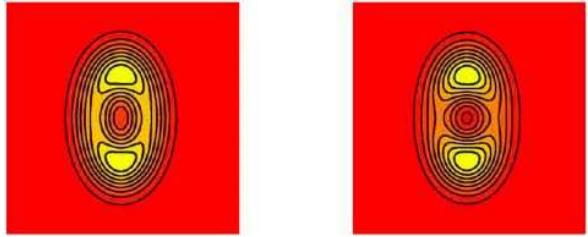}}
  \caption{\label{fig:two-body}(Color online) Two-point density
    $\langle \rho_t(\vect{r}_1, \vect{r}_2) \rangle_{|Q\rangle}$ with
    $\vect{r}_1 = 0$.  The left plot is for $Q = 0$ and the right for
    $Q = 1$.  Both plots have $\omega_y / \omega_x = 1/2$ and
    $\omega_c / \omega_x = 5$.}
\end{figure}

Physical differences in the qubits arise at the two-body level.  For
the two-particle density
\begin{equation}
  \label{eq:2BodyDensity}
  \hat{\rho}_t(\vect{r}_1, \vect{r}_2) = \frac{1}{2}
  \sum_{ij} \delta(\vect{r}_1 - \hat{\vect{r}}_i) \delta(\vect{r}_2 -
  \hat{\vect{r}}_j),
\end{equation}
we find
\begin{equation}
  \delta \rho = 2 F_{01} - F_{02} - F_{12},
\end{equation}
where $\delta\rho = \langle \rho_t \rangle_{|1\rangle} - \langle
\rho_t \rangle_{|0\rangle}$, and 
\begin{equation}
  F_{ij} = \text{Re} \left[
    \psi_i(\vect{r}_1) \psi_j(\vect{r}_2)
    \psi_i^*(\vect{r}_2) \psi_j^*(\vect{r}_1)\right].
\end{equation}
The two-point functions are shown in Fig.~\ref{fig:two-body} for
$\vect{r}_1 = 0$.

Because both the single-particle charge and spin properties for both
qubit states are identical, intra-qubit decoherence and dissipation
should be minimized.  We show below that, within the lowest-order Born
and Markov approximations, the $T_1$ and $T_2$ times are infinite for
a very large class of environmental models.  Before doing so, however,
we first show that a complete universal set of logic gates is
achievable in this system.

\section{Universal Quantum Logic Gates}
\label{sec:univ-quant-logic}

In the space defined by the qubit states in Eq.~(\ref{eq:qubits}), the
electronic Hamiltonian~(\ref{eq:ellipHamil}) can be written as a
pseudospin-1/2 particle in a pseudomagnetic field.  In this particular
case, we have\cite{Kyriakidis05:Coherent-rotations}
\begin{equation}
  \hat{H}_{\text{qubit}} = b_x\hat{\sigma}_x + b_z\hat{\sigma}_z + b_0
  \hat{\sigma}_0,
\end{equation}
with $\hat{\sigma}_x$, $\hat{\sigma}_z$ the Pauli spin matrices and
$\hat{\sigma}_0$ the identity matrix.  The pseudomagnetic field
components $b_x$, $b_z$, and $b_0$ are given by
\begin{subequations}
  \label{eq:pseudofield}
  \begin{gather}
    \label{eq:bx}
    b_x = \sqrt{3} \left( V_{0220} - V_{1221} \right) / 2,
    \\ \label{eq:bz}
    b_z = -V_{0110} + \left(V_{1221} + V_{0220}\right) / 2,
    \\ \label{eq:b0}
    b_0 = V_{0101} + V_{0202} + V_{1212},
  \end{gather}
\end{subequations}
where $V_{ijji}$ are exchange (and $V_{ijij}$ direct) matrix elements,
given in Eq.~(\ref{eq:VMatel}) with $n_i = n_j = 0,\ m_i = i$, and
$m_j = j$.  Explicit (exact, analytic) expressions of these are given
in Refs.~[\onlinecite{Kyriakidis05:Coherent-rotations,unpublished}].

The main point with regard to qubit rotations is that the fields in
Eq.~(\ref{eq:pseudofield}) have a different functional dependence on
the dimensionless ratios $r = \omega_y / \omega_x$ and $z = \omega_c /
\omega_x$.  Thus, adiabatically controlling either $r(t)$ or $z(t)$
can rotate qubits.\cite{adiabNote}  These ratios may be changed at
\emph{fixed} magnetic field ($\omega_c$) by altering the two
confinement frequencies $\omega_x$ and $\omega_y$ independently.

To perform an arbitrary computation, we require a universal set of
quantum logic gates which typically consists of both single and double
qubit operations.  We focus first on the single-qubit portion of this
universal set, followed by the two-qubit portion, the \textsc{cnot}
gate.

\subsection{Single Qubit Gates}
\label{sec:single-qubit-gates}

A universal set\cite{Nielsen00:Quantum-Computation} of quantum logic
gates is given by the \textsc{cnot} gate, which we discuss below, and
the single-qubit Hadamard gate $H$, $\pi / 8$ gate $T$, and phase gate
$S$.  These are each given by
\begin{equation}
  \label{eq:1bitgates}
  H = \frac{1}{\sqrt{2}}
  \begin{pmatrix} 1 & 1 \\ 1 & -1 \end{pmatrix},\quad
  T = \begin{pmatrix} 1 & 0 \\ 0 & \text{e}^{i\pi/4} \end{pmatrix},\quad
  S = \begin{pmatrix} 1 & 0 \\ 0 & i \end{pmatrix}.
\end{equation}
We expect that all SU(2) operations on qubits encoded as in
Eq.~(\ref{eq:qubits}) can be achieved asymptotically since two
non-parallel pseudofields are achievable with two different values of
$r$ and $z$.\cite{stepan04:univer.quant.comput}

In order to find explicit time-dependent parameters $r(t)$ and $z(t)$
for which the single-dot time evolution
\begin{equation}
  \label{expo2}
  \hat{U} = \widehat{\mathcal{T}} \exp \left( -i \int_0^T \! 
    \hat{H}_{\text{dot1}}(r(t),z(t)) \, dt  \right),
\end{equation}
equals the desired single-qubit operation, we adapt the minimization
method used in Refs.~[\onlinecite{Burkard99:Physical-optimization,
  Divincenzo00:Universal-quantum}].  ($\widehat{\mathcal{T}}$ is the
time-ordering operator.)  The time interval $[0,T]$ is divided into
$N$ discrete pieces during which the functions $r(t)$ and $z(t)$ are
set constant.  We are then left with an optimization problem with $3N$
variables $t_i$, $r_i$, and $z_i$ ($i=1,..,N$) where $t_i$ denotes the
length of the $i$th phase, $\sum_{i=1}^N t_i = T$; the parameters
$r_i$ and $z_i$ determine the values of $r(t)$ and $z(t)$ in the $i$th
phase.  We numerically minimize the function $f =
||U(\{t_i,r_i,z_i\}) - U_t||^2$ where $U(\{t_i,r_i,z_i\})$ is obtained
by exponentiation, Eq.~(\ref{expo2}), and $U_t$ is the desired target
single-qubit operation, Eq.~(\ref{eq:1bitgates}).

We have found numerical solutions involving $N=1$, 3, and 5 steps for
the $H$, $T$, and $S$ gates respectively.  Explicit sequences are
shown in Table~\ref{tab:1bitgates},\cite{accuracy} where the time
pulse duration is expressed in terms of the dimensionless parameter
$\tau = t/t_0$ with $t_0 = (2\pi/\omega_x) [8 \pi^2 \hbar
\omega_x/\text{Ry}]^{1/2}$.  Here, $\text{Ry} = m^* e^4/ (2 \epsilon^2
\hbar^2)$ is the effective Rydberg energy, $m^*$ is the effective
mass, and $\epsilon$ the dielectric constant.  For GaAs, $\text{Ry}
\approx 5.93$~meV and $t_0\approx 2.5\,{\rm ps}$ for
$\omega_x=2.5\,{\rm meV}$.
\begin{table}
  \caption{\label{tab:1bitgates}Pulse sequences for
    one-bit logic gates.  The
    dimensionless parameters $\tau$, $z$, and $r$ can be tuned through
    the time $t$ and any \emph{two} of $\omega_y$, $\omega_x$, and
    $B_z$.}
  \begin{ruledtabular}
    \begin{tabular}{ccccccccc}
      \multicolumn{3}{c}{Hadamard gate} &
      \multicolumn{3}{c}{$\pi / 8$ gate} &
      \multicolumn{3}{c}{Phase gate}\\
      $2\pi \tau$& $z$  & $r$  &$2\pi \tau$& $z$  & $r$  &$2\pi\tau$& $z$  & $r$  \\ \hline
      1.470 & 0.376& 0.158& 1.859& 4.828& 0.022& 2.092& 0.249& 0.121\\
      {}    &      &      & 3.674& 0.102& 0.936& 1.512& 2.803& 0.996\\
      {}    &      &      & 2.443& 1.093& 0.051& 2.123& 2.586& 0.012\\
      {}    &      &      &      &      &      & 2.280& 0.124& 0.916\\
      {}    &      &      &      &      &      & 1.992& 0.224& 0.139
    \end{tabular}
  \end{ruledtabular}
\end{table}
Note that the sequences shown in Table~\ref{tab:1bitgates} are not
optimized for experimental efficiency, but merely demonstrate that
solutions for a universal set do indeed exist.  We have found many
more solutions (none shorter) for each of the one-bit gates, including
solutions at fixed $z$.\cite{unpublished}

\subsection{Coupled Dots}
\label{sec:coupled-dots}

\begin{figure}
  \resizebox{6cm}{!}{\includegraphics*{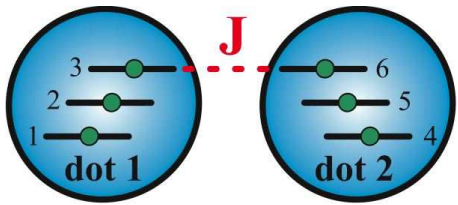}}
  \caption{\label{fig:coupling}(Color online) Schematic of our minimal
    coupling model.  As the interdot barrier is lowered, the orbitals
    highest in energy and with the greatest overlap will be the first
    to couple.}
\end{figure}

To consider two-qubit gates, we now include the interdot coupling term
$\hat{H}_{\text{coupl}}$.  We consider a minimal model valid in the
limit of weak coupling.  Of the three orbitals we are considering, the
$|nm\rangle = |02\rangle$ orbital is both highest in energy and
closest to the edge of the dot.  Thus, as the inter-dot barrier is
lowered, the respective $|02\rangle$ orbitals in each dot will be the
first to couple.  This is schematically depicted in
Fig.~\ref{fig:coupling}.  Our minimal model considers the coupling
only between these two orbitals.  This leads to a Heisenberg
form,\cite{Loss98:Quantum-computation, Burkard99:Coupled-quantum}
\begin{equation}
  \hat{H}_{\text{coupl}} = J \hat{\vect{S}}_{\ell} \cdot
  \hat{\vect{S}}_r,
\end{equation}
where $\hat{\vect{S}} = \sum_{ss'} c_{02s}^\dagger \vect{\sigma}_{ss'}
c_{02s'}$ is the spin operator of the $|02\rangle$ orbital, and the
indices $\ell,\ r$ denote the left and right dot respectively.

\subsection{Double Qubit Gate}
\label{sec:cnot-gate}

A two-qubit system is formed from the direct product $|Q\rangle
\otimes |Q'\rangle$, of the states in Eq.~(\ref{eq:qubits}), forming a
four-dimensional computational space.  The states $|Q\rangle \otimes
|Q'\rangle$ are, in fact, $S = 1,\ S_z = -1$ eigenstates.
Unfortunately, the spin subspace and the computational subspace are
not identical; the six spin (three for each dot) $S = 1,\ S_z = -1$
subspace is nine-dimensional, four of which constitute our $|Q\rangle
\otimes |Q'\rangle$ computational space.  Thus, our implementation of
\textsc{cnot}, as that in
Ref.~[\onlinecite{Divincenzo00:Universal-quantum}], involves transient
excursions outside the computational space; nevertheless, our
sequences are designed such that the final gate operation is unitary
and returns to the four-dimensional computational space.  We require
the final state to be such that the \textsc{cnot} truth table be
satisfied, up to single-qubit operations.\cite{Makhlin02:Invariants}
An explicit implementation is given in Table~\ref{tab:cnot}.
\begin{table}
  \caption{\label{tab:cnot}\textsc{Cnot} implementation with an
    always-on intradot exchange interaction.  Subscripts denote
    individual qubits, $\mathcal{J}$ the interdot exchange, and $\tau$
    the pulse duration.}
  \begin{ruledtabular}
    \begin{tabular}{cccccc}
      $2\pi \tau$ & $\mathcal{J}$ & $z_1$ & $r_1$ & $z_2$ & $r_2$\\ \hline
      1.227 & 2.133 & 0.846 & 0.630 & 3.280 & 0.398\\
      3.821 & 0.615 & 1.860 & 0.067 & 0.663 & 0.308\\
      2.766 & 4.094 & 0.418 & 0.767 & 3.897 & 0.340\\
      1.167 & 3.540 & 0.017 & 0.298 & 0.852 & 0.952\\
      1.591 & 3.242 & 1.695 & 0.370 & 2.362 & 0.237\\
      2.148 & 3.031 & 2.177 & 0.559 & 2.648 & 0.354\\
      1.560 & 1.714 & 3.091 & 0.077 & 4.812 & 0.083\\
      2.255 & 1.889 & 1.536 & 0.222 & 2.032 & 0.645\\
      1.981 & 3.796 &21.501 & 0.453 &11.516 & 0.157
    \end{tabular}
  \end{ruledtabular}
\end{table}
Six parameters are required to describe a pulse: the pulse duration
$\tau$, two ($r$, $z$) parameters per dot to describe each qubit, and
a dimensionless exchange coupling $\mathcal{J} = J t \hbar / \tau$
describing the coupling.  As shown in Table~\ref{tab:cnot}, a nine
step solution is the smallest we have been able to find.\cite{cnot}
(If it were possible to turn off intradot exchange, then a three-pulse
\textsc{cnot} is achievable.\cite{unpublished})

\section{Decoherence and Dissipation}
\label{sec:decoh-diss}

Environmental influences can be of two distinct types: Slow variations
in the electromagnetic environment merely lead to adiabatic changes in
the pseudofield ${\bf b}$ and thus to unitary errors that typically
average out over the length of a pulse.\cite{adiabNote} We look first
to fast, non-adiabatic environmental influences that can lead to
non-unitary errors---i.e., decoherence---followed by a discussion on
adiabatic influences which lead to gate errors.

\subsection{Nonadiabatic Influences}
\label{sec:nonad-infl}

Assuming the environment does not change the number of particles on
the dot, and that the bath couples only to single particles in the
dot, then a general model of system-bath coupling is given by
\begin{equation}
  \label{eq:hSB}
  \hat{H}_{\text{SB}} = \sum \hat{B}^{n'm's'}_{nms}\, 
  c^{\dagger}_{n'm's'} c_{nms},
\end{equation}
where the sum is over all repeated indices.  $\hat{B}^{n'm's'}_{nms}$
is a set of arbitrary operators which describe the reservoir and all
relevant coupling constants.

At time $t$, the full state-vector of the system $|\Phi(t)\rangle =
\sum_Q |Q\rangle \otimes |\chi_Q(t)\rangle$ ($Q = 0,\ 1$) is a
non-separable state, where the states $|\chi_Q(t)\rangle$ are
reservoir states including all time-dependent coefficients.  Matrix
elements of Eq.~(\ref{eq:hSB}),
\begin{equation}
  \label{eq:hSBMatElem}
  H_{\text{SB}}^{Q'Q} = \sum \langle Q'|c^{\dagger}_{n'm's'}
  c_{nms}|Q\rangle\, A_{nms}^{n'm's'}(\chi',\chi),
\end{equation}
where $ A_{nms}^{n'm's'}(\chi',\chi) =
\langle\chi'|\hat{B}_{nms}^{n'm's'}|\chi\rangle$ are straightforwardly
calculated.\cite{unpublished}

Within the Born-Markov approximation,
and using the definitions Eq.~(\ref{eq:qubits}),
the relaxation $T_1$ and dephasing $T_{\varphi}$ times are given
by\cite{Abragam61,khaet03:elect.spin.evolut} ($1/T_2 = 1/(2T_1) + 1/T_{\varphi}$)
\begin{subequations}
  \label{eq:born-markov}
  \begin{gather}
    \label{eq:T1}
    \frac{1}{T_1} \sim \left|H_{\text{SB}}^{10}\right|^2 = 
    \frac{1}{12} \left|\delta h_0 - \delta h_1\right|^2,
    \\
    \label{eq:T2}
    \frac{1}{T_{\varphi}} \sim \left|H_{\text{SB}}^{00} -
      H_{\text{SB}}^{11}\right|^2 = \frac{1}{9} \left| 2 \delta h_2 -
      (\delta h_1 + \delta h_0) \right|^2,
  \end{gather}
\end{subequations}
where $\delta h_m = A_{0m\uparrow}^{0m\uparrow} -
A_{0m\downarrow}^{0m\downarrow}$.  To the extent that the Born-Markov
approximation is valid,\cite{DiVincenzo05:Rigorous-Born}
Eq.~(\ref{eq:born-markov}) states that relaxation and dephasing within
the computational space are negligible to leading order for all
environmental couplings which are either purely charge or purely spin
in character.  The former has $\hat{B}^{m'n's'}_{mns} =
\delta_{ss'}\hat{B}^{m'n'}_{mn}$ in Eq.~(\ref{eq:hSB}) and
consequently $\delta h_m = 0$, whereas the latter has
$\hat{B}^{m'n's'}_{mns} = \delta_{m'm} \delta_{n'n} \hat{B}^{s'}_{s}$
and consequently $\delta h_0 = \delta h_1 = \delta h_2 \neq 0$.  For
both these cases, dephasing and relaxation vanish within the
Born-Markov approximation.  (Neither of these is applicable for
hyperfine environments which depend on \emph{both} spin and charge.)

\subsection{Adiabatic Influences}
\label{sec:adiabatic-influences}

Regarding adiabatic (unitary) influences, which do not cause
decoherence, these can be minimized by choosing settings for the
confinement potential such that $d\vect{b}/dr$ and $d\vect{b}/dz$ are
small in magnitude.  This is the case, for example, for the values $r
\approx 0.8$ and $z \rightarrow 0$.  For these values of the
confinement, we find $|d\vect{b}/dz| \rightarrow 0$ and
$|d\vect{b}/dr| \sim e^2 / (16 \sqrt{2 \pi} \lambda)$, where
$\lambda^2 = \hbar / (m \omega_x)$.  For GaAs material parameters,
with $\omega_x = 1$~meV, this gives $|d\vect{b}/dr| \approx
85\,\mathrm{\mu}$eV---at least an order of magnitude smaller than
typical pseudofield magnitudes.  With these, we can estimate
corrections to the adiabatic limit\cite{messiah99:quant_mechan}
expressed as a leakage time given by $T_{\mathrm{leak}} \sim E^4 /
(\hbar^2 \dot{r}^3 |d\vect{b}/dr|^2)$, where $E \sim 1$~meV is the
excitation energy to states outside the qubit space, and $\dot{r} =
dr/dt \sim 10$~GHz is the rate of typical gate operation.  For these
parameter values, we find $T_{\mathrm{leak}} \sim 200\ \mu s$, and a
leakage probability of only $P_{\mathrm{leak}} \sim 10^{-7}$.

\section{Initialization and Measurement}
\label{sec:init-meas}

With regard to state initialization, we note that if the qubit states
are the lowest-energy states, then preparation becomes merely a matter
of thermalization.  In this case, potentials other than elliptic may
well prove useful.\cite{ellipNote}  Ideally, a system where the two
qubit states are the two lowest-energy states would be beneficial not
only for state preparation, but also for a more general deterrent to
dissipation, especially relaxation to states outside the computational
basis.

Finally, with regard to measurement, we note that the two states in
Eq.~(\ref{eq:qubits}) are not degenerate.  Thus, a destructive
measurement is possible by detecting whether a fourth electron
resonantly tunnels onto the (three-particle) dot; similarly to the
single-shot readout of individual quantum dot
spin,\cite{elzer04:singl.shot.read} the gates may be pulsed such that
an additional electron can tunnel onto the dot only if it is in the
higher-energy qubit state.  In fact, since the tunnel barriers as well
as the confinement itself is determined (and controlled) by the
applied electrostatic potential, the universal set of gates described
above as well as detection may be accomplished using
already-existing\cite{elzer04:singl.shot.read,
  Petta05:Coherent-Manipulation-of-Coupled} experimental techniques.

\begin{acknowledgments}
  JK acknowledges support from CFI and NSERC of Canada.  GB
  acknowledges funding from the Swiss SNF and NCCR Nanoscience.
\end{acknowledgments}

% \bibliography{dotqcbib,0D,misc}
%\bibliography{dotqc} % Self contained bibtex file

\begin{thebibliography}{25}
\bibitem[{ell()}]{ellipNote}
\bibinfo{note}{Elliptic confinement is sufficient but not necessary for our
  scheme to work. In particular, confinements of less symmetry will certainly
  also work---purely circular dots will not. The potential essentially should
  be describable by at least two independent parameters. Elliptic confinement
  is the simplest such potential.}
\bibitem[{adi()}]{adiabNote}
\bibinfo{note}{Changes in the external charge environment are adiabatic if they
  occur on time scales longer than $\hbar/(e^2/\sqrt{\hbar/m\omega_0})$.}
\bibitem[{cno()}]{cnot}
\bibinfo{note}{We have found 4 9-step solutions and 25 10-step solutions.}
\end{thebibliography}

\end{document}